\documentclass[%
 aip,
 jmp,%
 amsmath,amssymb,
preprint,%
]{revtex4-1}

\usepackage[english]{babel}
\usepackage{graphicx}
\usepackage{dcolumn}
\usepackage{bm}

\usepackage{verbatim}

\graphicspath{ {final_plots/} }

\begin{document}

\preprint{}

\title{Position-dependent and pair diffusivity profiles from steady-state solutions of color reaction-counterdiffusion problems}

\author{James Carmer}
\affiliation{McKetta Department of Chemical Engineering, The University of Texas at Austin, Austin, TX 78712.
}
\author{Frank van Swol}%
\affiliation{
Sandia National Laboratories, Department 1814, P.O. Box 5800, Albuquerque, NM 87185.
}%

\author{Thomas M. Truskett$^{1,}$}
 \email{truskett@che.utexas.edu}

\date{\today}
             

\maketitle

Hydrodynamic theory predicts that diffusivity of a particle near an interface in a viscous, continuum solvent depends on spatial position.\cite{Happel1973LowReynoldsNumber} If the surrounding fluid is non-continuum--e.g., comprises particles of comparable size to the tagged particle--then the diffusivity profile is strongly influenced by the medium's static structure.\cite{Mittal2008LayeringandPosition,mittal:034110,netz2011}
Such position-dependent dynamics, while challenging to characterize,\cite{Mittal2008LayeringandPosition,doi:10.1021/jp0375057} are critical for understanding and modeling kinetics in colloidal and interfacial fluid systems.

Recently, a novel stochastic approach was introduced~\cite{Mittal2008LayeringandPosition,mittal:034110} for estimating diffusivity profiles of inhomogeneous fluids consistent with time- and position-dependent particle displacement data [obtained from, e.g., molecular dynamics (MD) simulations or confocal microscopy experiments] using Bayesian inference or likelihood maximization.
In this Note, we show how a simple and physically intuitive particle labeling strategy can be used to obtain the same profiles via the steady-state solution of a color reaction-counterdiffusion problem. 

First, we consider a dense fluid of $N$ hard spheres (HS) of diameter $\sigma$ and mass $m$ confined to a slit pore of width $H$ by parallel hard walls of area $A$, i.e., with nominal packing fraction $\phi=N \pi \sigma^3 /(6AH)$. Below, we report quantities for this system implicitly nondimensionalized by appropriate combinations of characteristic scales for length ($\sigma$) and energy ($\beta^{-1})$, where $\beta^{-1}=k_{\text{B}} T$, $k_{\text{B}}$ is the Boltzmann constant, and $T$ is temperature.

\begin{figure}[ht]
  \centering
  \includegraphics[width=0.45\textwidth]{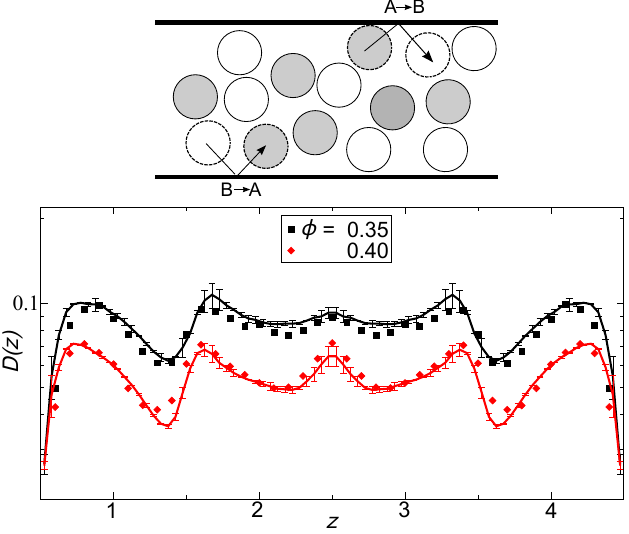}
  \caption{Top: Labeled particles changing color upon a ``reactive'' wall collision. Bottom: Comparison between HS diffusivity profiles $D(z)$ estimated from the steady-state color analysis (curves with bars indicating 95\% confidence intervals) and a Bayesian inference technique (points)\cite{Mittal2008LayeringandPosition} for pore width $H = 5$.}
  \label{fig:schematic_result2}
\end{figure}

The Markovian propagator $G(z;\delta t| z',0)$ for single-particle displacements in such systems--over intermediate to long lag times $\delta t$--obeys the Smoluchowski (Fokker-Planck) equation:\cite{Mittal2008LayeringandPosition}
  \begin{equation}
  \label{eq:smo_diff}
    \frac{\partial G}{\partial t} = \frac{\partial}{\partial z} \left\{D(z) e^{-F(z)} \frac{\partial}{\partial z}\left[ e^{F(z)} G\right] \right\}
   \end{equation}
where $D(z)$ is diffusivity, $F(z) = -\ln \rho(z)$, and $\rho(z)$ is number density.

We assign a color (A or B) to each particle and create opposing ``color reaction" surfaces at the walls (Fig~\ref{fig:schematic_result2}) where particles of a specific color (depending on the wall) can transform to the other color with probability $p_r$. This reaction-counterdiffusion process, modeled by Eq.~\ref{eq:smo_diff}, evolves to a steady state with the flux $j_i$ of particle type $i$ expressed as $j_i = - D(z) \rho (z) {d}x_i/{d}z$, where $x_i$ is the mole fraction.
We rearrange to get $D(z)$ in terms of quantities measurable via particle tracking:
  \begin{equation}
	 \label{eq:diff_z}
     D(z) = \frac{-j_i}{\rho (z) \: \mathrm{d}x_i / \mathrm{d}z}
  \end{equation}
To compute $D(z)$ via eq.~\ref{eq:diff_z}, we perform simulations using discontinuous MD (see, e.g., \cite{Rapaport2004TheArtOf}) in the canonical ensemble with $2000 \le N \le 4000$, depending on $\phi$. 
The $z$ coordinate of the pore is divided into bins of width $1/50$, and $\rho(z)$ and $x_i(z)$ are calculated from average particle numbers and color compositions in each bin. Taking the first derivative of a local cubic fit to the composition profile gives $\mathrm{d}x_i / \mathrm{d}z$.
The flux $j_i$ is measured from the steady-state reaction rate at the walls.

The diffusivity profile $D(z)$ obtained from the color labeling approach with $p_r=0.01$ agrees with the earlier Bayesian analysis\cite{Mittal2008LayeringandPosition}, as can be seen in the bottom panel of Fig.~\ref{fig:schematic_result2}. In contrast to the color method, there is a choice to make about which intermediate lag times $\delta t$ to consider in the Bayesian inference approach\cite{Mittal2008LayeringandPosition}; motion is not diffusive at short times, and information about position dependence of particle dynamics is washed out at long times.   
\begin{figure}[ht]
  \centering
  \includegraphics[width=0.4\textwidth]{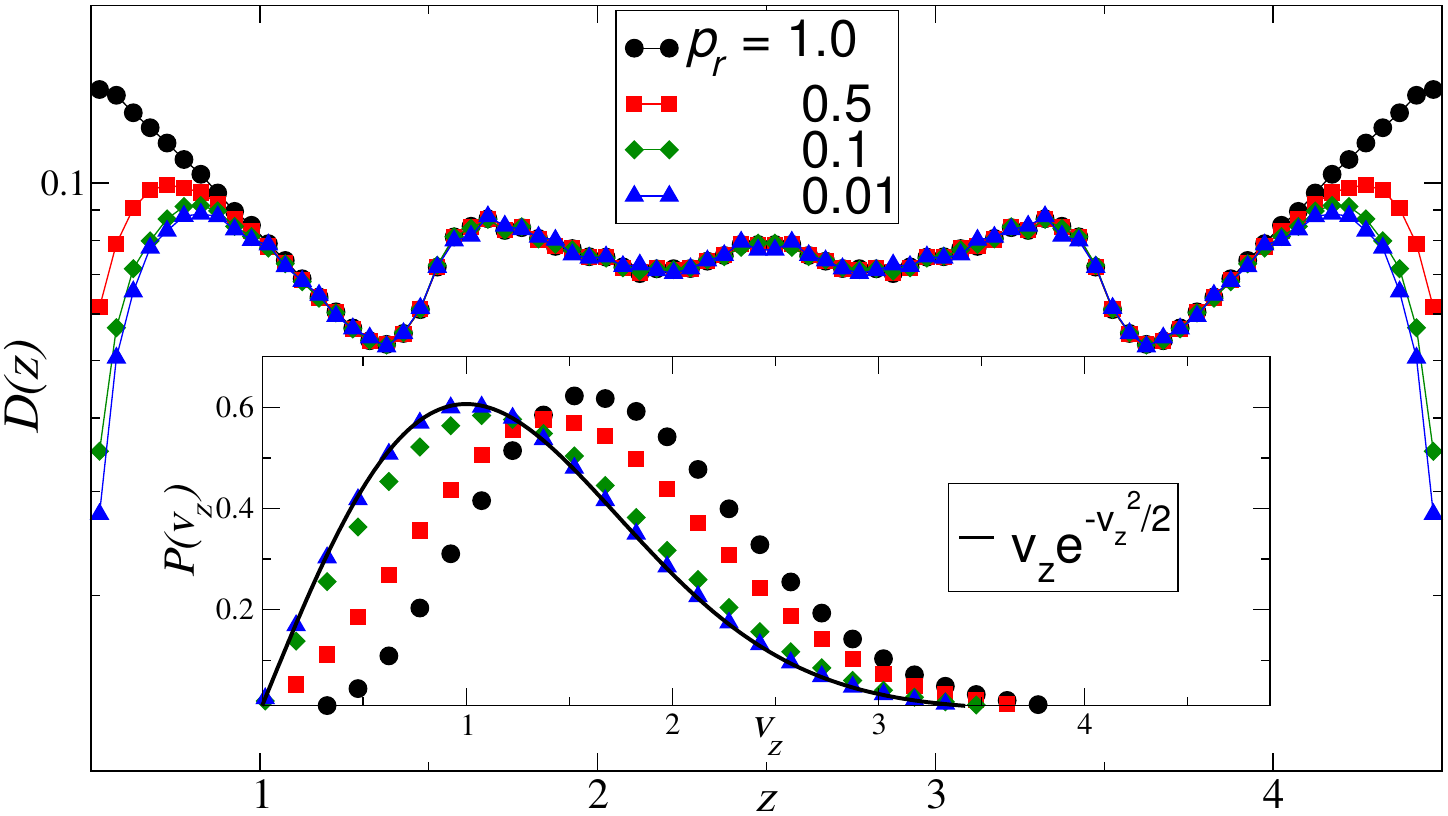}
  \caption{The effect of reaction probability $p_r$ on estimated confined HS $D(z)$ at $\phi = 0.35$ and $H = 5$ from the steady-state color analysis. Inset: The corresponding probability distributions of velocities normal to the boundary for reacting particles.}
  \label{fig:probability_dvz}
\end{figure}

Fig.~\ref{fig:probability_dvz} shows the effect of $p_r$ on $D(z)$ computed from color diffusion. If $p_r >0.1$, then we see that diffusivities exhibit an unwanted $p_r$ dependence near the boundaries and no longer match the data from the Bayesian analysis. This is understood by looking at the probability distribution of normal velocities after a reactive collision with a boundary. If $p_r$ is chosen to be too high, then the kinetics of the fictitious surface reaction become important and successfully reacted particles rebound with velocities higher than expected based on the equilibrium Maxwell-Boltzmann distribution. The kinetic bias toward faster rebounding particles--and artificially high computed diffusivities near the wall--is removed as $p_r$ is lowered and the equilibrium velocity distribution of reacted particles is recovered, which can provide practical guidance in choosing $p_r$.

\begin{figure}[ht]
  \centering
  \includegraphics[width=0.45\textwidth]{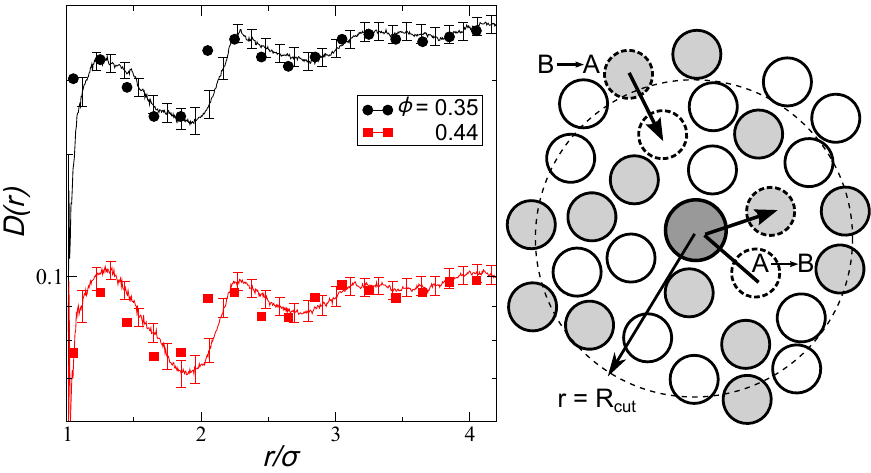}
  \caption{Left: Comparison between pair diffusivity profiles $D(r)$ of the HS fluid estimated from the steady-state color analysis (curves) and a recent Bayesian inference technique (points) \cite{doi:10.1021/jp0375057}. Right: Schematic for calculating $D(r)$ using the color diffusion method. Labeled particles changing color upon successful ``reactive'' boundary events, colliding with the central particle or crossing an imaginary reaction surface (in the direction pictured) a distance $R_{cut}$ from the central particle.}
  \label{fig:tracer}
\end{figure}

The same approach can be applied to systems for which the ``reaction surface'' is not a physical boundary. Fig.~\ref{fig:tracer} illustrates how it can be recast to determine the normal pair diffusivity $D(r)$ of particles in a bulk, isotropic fluid. A particles can react to form B particles upon colliding with a central particle, and B particles can react to form A particles when they enter a region defined by an imaginary boundary placed at $R_{cut}$ (a distance larger than any relevant static or dynamic correlation length in the system). As before, diffusivity is related to steady-state color fluxes and compositions:   
\begin{equation}
  \label{eq:diff_r}
  D(r) = -\frac{j_i(r)}{\rho (r) \: \left(\mathrm{d}x_i / \mathrm{d}r - 2 x_i / r \right)}
\end{equation}
In Fig.~\ref{fig:tracer}, we compare $D(r)$ of the HS fluid estimated from Eq.~\ref{eq:diff_r} and from the Bayesian inference technique\cite{doi:10.1021/jp0375057}. For the former, the MD simulations have $N=4000$ particles, $R_{cut}=8$, and, as in Fig.~\ref{fig:schematic_result2}, $p_r=0.01$ at both reaction surfaces. As can be seen, there is again good agreement between the two approaches.

Color reaction-diffusion processes in inhomogeneous HS fluids represent a rich class of problems, and a detailed study~\cite{Frank} examining their properties will soon be presented elsewhere.   
In the future, it would also be interesting to explore use of color labeling to study systems with soft particles or boundaries, as well as to compare it with another Fokker-Planck based approach~\cite{netz2011} that estimates diffusivity profiles from mean-first-passage-time data. 

TMT acknowledges support from the Welch Foundation (F-1696), the Gulf of Mexico Research Intitiative, and the Texas Advanced Computing Center (TACC) at The University of Texas at Austin. FVS acknowledges support by the US DOE, Office of Basic Energy Sciences, Division of Materials Sciences and Engineering and Sandia's LDRD program. Sandia National Laboratories is a multi-program laboratory managed and operated by Sandia Corporation, a wholly owned subsidiary of Lockheed Martin Corporation, for the U.S. Department of Energy's National Nuclear Security Administration under contract DE-AC04-94AL85000.







\end{document}